\documentclass{article}
\usepackage{spconf,amsmath,graphicx, booktabs, graphicx, hyperref,textcomp}

\makeatletter

\renewcommand{\section}{\@startsection
  {section}
  {}
  {}
  {-\bigskipamount}%
  {\bigskipamount}%
  {}}%

\renewcommand\section{\@startsection 
{section}
{1}
{\z@}%
{-2.5ex \@plus -1ex \@minus -.2ex}%
{2.3ex \@plus.2ex}%
{\normalfont\Large\bfseries}
}

\renewcommand\subsection{\@startsection
{subsection}
{2}
{\z@}
{-2.25ex\@plus -1ex \@minus -.2ex}%
{1.5ex \@plus .2ex}%
{\normalfont\large\bfseries}}

\makeatother

\def\math#1{\ifmmode#1\else$#1$\fi}

\newcount\exponent
\def\e{\afterassignment\ee\exponent=}
\def\ee{\math{10^{\the\exponent}}}

\newcommand{\textapprox}{\raisebox{0.5ex}{\texttildelow}}

\title{Fine-tuning wav2vec2 for speaker recognition}
\name{Nik Vaessen, David A. van Leeuwen
}
\address{Institute for Computing and Information Sciences, Radboud University, Nijmegen, The Netherlands\\\texttt{\{nvaessen,dvanleeuwen\}@science.ru.nl}}

\begin{document}
%
\maketitle
\begin{abstract}

This paper explores applying the wav2vec2 framework to speaker recognition instead of speech recognition. We study the effectiveness of the pre-trained weights on the speaker recognition task, and how to pool the wav2vec2 output sequence into a fixed-length speaker embedding. To adapt the framework to speaker recognition, we propose a single-utterance classification variant with cross-entropy or additive angular softmax loss, and an utterance-pair classification variant with BCE loss. Our best performing variant achieves a 1.88\% EER on the extended voxceleb1 test set compared to 1.69\% EER with an ECAPA-TDNN baseline. Code is available at \href{https://github.com/nikvaessen/w2v2-speaker}{\texttt{github.com/nikvaessen/w2v2-speaker}}.
\end{abstract}
\section{Introduction}
\label{sec:intro}

In the field of natural language processing (NLP) it has become standard to fine-tune self-supervised pre-trained models, such as BERT \cite{devlin2018bert}, XLNet \cite{yang2019xlnet}, and T5 \cite{raffel2019exploring}, on a wide variety of NLP tasks. Recently, this framework of pre-training and fine-tuning has also been successfully used in automatic speech recognition with wav2vec2 \cite{NEURIPS2020_92d1e1eb}. The aim of this study is to explore the feasibility of fine-tuning the wav2vec2 pre-trained network on a different task than speech recognition, namely \emph{speaker} recognition. 

The BERT and wav2vec2 network have commonalities in their design. Both have stacks of transformer layers and they use self-supervised, contrastive pre-training with masked input. However, they differ in three major aspects: 1) the input tokens to the encoder in wav2vec2 are raw audio processed by a CNN instead of WordPiece embeddings, 2) wav2vec2 uses relative positional embeddings computed by a CNN instead of sinusoidal positional embeddings, 3) there is no class token and equivalent next-sentence prediction task in the pre-training procedure of wav2vec2.

The class token in BERT is used when fine-tuning on sentence-pair classification tasks, e.g., entailment, and for single-sentence classification tasks, e.g., sentiment analysis. It is not used for single sentence tagging tasks, e.g., named entity recognition. In this respect, speech recognition is analogous to sentence tagging where each output token can represent a phone or letter. Hence, speech recognition does not require a class token.
In contrast, speaker recognition corresponds to either sentence-pair classification, or single-sentence classification, and might therefore benefit from a class token which summarizes the whole input sequence. 

Our work focuses on the following questions. First and foremost, we want to find out whether the pre-trained wav2vec2 weights are an effective initialization for speaker recognition. We further want to explore if wav2vec2 can be adapted to speaker recognition without a class token and next-sentence prediction task. One solution is to pool the variable-length sequence of wav2vec2 embeddings into a fixed-size speaker embedding. This raises the questions if pooling is an effective replacement for a class token, and if so, which pooling method is most suitable.

\section{Related work}
\label{sec:related}

Wav2vec2 has already been applied to a variety of speech-related tasks. 
In \cite{fan21_interspeech} the network is used for speaker recognition and language identification in both a single and multi-task learning setting. 
The authors of \cite{tjandra2021improved} show good performance for language identification with 25 languages but modify wav2vec2 to use log-mel spectogram input 
instead of raw waveforms.
In \cite{pepino21_interspeech} the (frozen) wav2vec2 embeddings are input to a learnable downstream model for carrying out emotion recognition.
Meanwhile \cite{yuan2021role} manages to fine-tune the wav2vec2 model itself on emotion recognition with CTC loss by using emotion-labeled phonetic units. 
The LeBenchmark \cite{evain21_interspeech} uses the wav2vec2 models as a baseline and encapsulates speech recognition, spoken language understanding, emotion recognition and speech translation in a single benchmark.  
 


\section{Methodology}
\label{sec:methodology}


\subsection{The wav2vec2 architecture}

\begin{figure}
    \centering
    \includegraphics{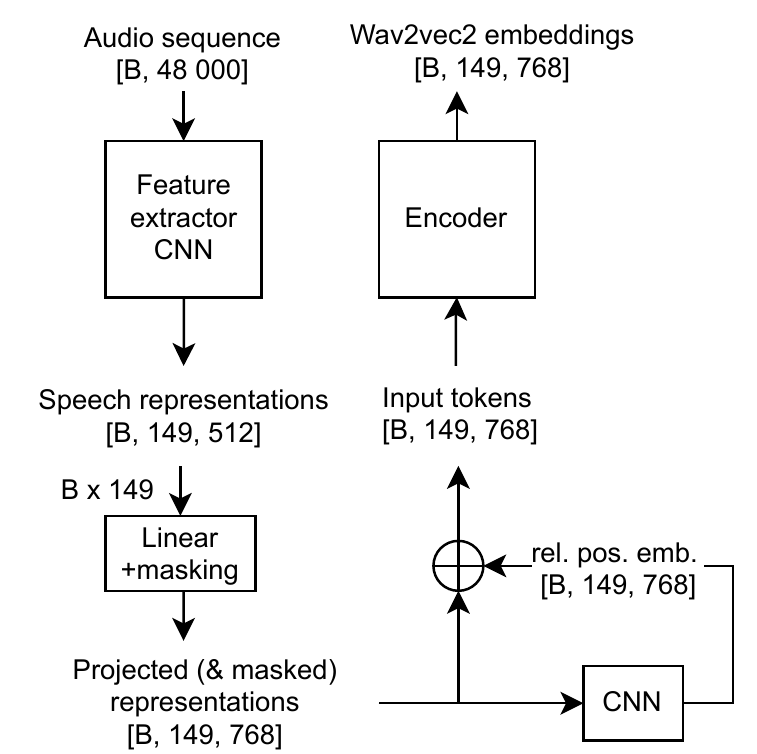}
    \caption{\fontsize{9}{11}\selectfont Overview of the Wav2vec2 architecture. Shapes are specified for a batch of $B$ audio samples with a length of 3 seconds.}  
    \label{fig:wav2vec2_architecture}
\end{figure}

The wav2vec2 framework \cite{NEURIPS2020_92d1e1eb} applies the concept of self-supervised pre-training with transformers to automatic speech recognition. In Fig.~\ref{fig:wav2vec2_architecture} we show a general overview of the network architecture during fine-tuning. The next subsections summarize each component.

\subsubsection{Feature extraction}

The first step is to encode a raw audio waveform (normalized to zero mean and unit variance) into learned representations with a discrete time unit. The \textit{feature extractor} consists of 7 consecutive 1-dimensional convolutions with 512 channels and respective kernel sizes of (10, 3, 3, 3, 3, 2, 2) and stride (5, 2, 2, 2, 2, 2, 2). The output of the first convolutional layer is group normalized \cite{wu2018group} such that each of the 512 channel sequences has zero mean and unit variance before GELU activation \cite{hendrycks2016gaussian} is applied. The other convolutional layers do not have any normalization layers and their output is directly activated with GELU. The output of the feature extractor is an encoded vector sequence with dimensionality 512. Each vector has a receptive field of 20\,ms which is similar to the window sizes in spectral-based representations. 

\subsubsection{Projection, SpecAugment \& positional embedding}

After the feature extraction each encoded vector representation in the sequence is independently normalized to zero mean and unit variance and projected into 768 dimensions by a single, shared fully-connected layer called the \textit{feature projector}. On all projections dropout is applied (but no activation). Then, masking is applied over the whole sequence analogous to  SpecAugment \cite{park19e_interspeech}; 0 or more random sets of consecutive vectors (masking in time domain) as well as 0 or more random sets of consecutive channels (masking in ``frequency'' domain) have their values blanked to 0. This masked projected sequence is then convolved by a single layer with a kernel size of 128, a stride of 1, padding of 64 and 16 groups followed by GELU activation in order to create a \textit{relative positional embedding} for each projected representation. This relative positional embedding is summed with the original input of the convolution, which changes the receptive field from 20\,ms to 2.5\,s. As a final step each vector is independently normalized with LayerNorm and dropout is applied again.

\subsubsection{Transformer}

The masked and projected sequence with both local and positional information is fed through an \textit{encoder} with 12 consecutive transformer layers. Each transformer layer consists of a residual 12-headed self-attention module and a residual 2-layer feed forward network with respectively 3072 and 768 units. LayerDrop \cite{huang2016deep,fan2019reducing} is applied such that each transformer layer is potentially skipped. The final output sequence, with each representation potentially having both local and global information due to self-attention, is used in a downstream task.

\subsection{Three wav2vec2 variants for speaker recognition}

The original wav2vec2 framework fine-tunes on speech recognition by independently labeling each wav2vec2 output embedding with a shared fully-connected layer, and optimizes with CTC loss \cite{graves2006connectionist}. 
We propose two adaptions to this design for the speaker recognition task, inspired by BERTs~\cite{devlin2018bert} single-sentence and sentence-pair classification setup.




\subsubsection{Speaker recognition as single-utterance classification}

The current paradigm in speaker recognition with deep neural networks is to train models with a classification-based approach~\cite{snyder2018x, desplanques20_interspeech}. To mimic this architectural paradigm two modifications to wav2vec2 are made. First, the sequence of wav2vec2 embeddings is reduced to a single embedding during training (see subsection \ref{sec:meth_pool}). Secondly, we add a fully connected layer which uses the pooled embedding to classify each speaker in the training data with cross-entropy (CE) or angular additive softmax (AAM) \cite{Deng_2019_CVPR,liu19f_interspeech} loss. A trial is evaluated with the cosine similarity between two pooled embeddings. We refer to these variants as \textit{w2v2-ce} and \textit{w2v2-aam}.

\subsubsection{Speaker recognition as utterance-pair classification}

The second approach directly computes a similarity score. The two audio segments of a speaker recognition trial are first processed independently up to the encoder part of the network. Then, the two sequences of input tokens are concatenated, accompanied by special tokens at the embedding level: A start (all $+1$), separator (all $-1$), and end (all $-1$) token. The first wav2vec2 embedding in the output sequence (corresponding to the start token) is used as input to a logistic regression with one dense layer. The singular output is the logit for the BCE loss and the score for evaluating a trial. During training a batch consists of 8 speakers, 4 utterances per speaker, and 16 same and different speaker-pairs. We refer to this third variant as \textit{w2v2-bce}.

\subsection{Pooling methods} \label{sec:meth_pool}

We propose several pooling methods to reduce the variable-length sequence of wav2vec2 embeddings to a fixed-size speaker embedding. We first consider the standard statistical pooling methods \textit{mean}, \textit{max}, \textit{mean\&std} and \textit{quantile}. They aggregate each dimension over the time axis. The \textit{mean\&std} variant doubles the embedding dimensionality while \textit{quantile} pooling expands each dimension five-fold with quantiles (0, 0.25, 0.5, 0.75, 1). We also assess taking the \textit{first}, \textit{middle} or \textit{last} embedding of the sequence as a ``pooling'' strategy as well as \textit{random}ly selecting the index. Lastly we consider inserting a ``start'' token (all values $+1$)  before the input sequence of the encoder and then selecting the first output token as the speaker embedding. This is termed \textit{first\&cls}. Unlike BERT our ``start'' token token does not have a meaningful prior.

\section{Experiments}
\label{sec:exp}

\subsection{Data}

All experiments are conducted on the VoxCeleb datasets \cite{Nagrani17, Chung18b}, which consist of interviews of celebrities extracted from YouTube. The VoxCeleb2 dev set, which contains \textapprox 1.1\,M audio files over 6\,k speakers, is used for training. A validation set is created based on \textapprox2\,\% of the dev set data which includes all speakers but does not overlap in recordings. For this validation set a random trial set of 5\,k same and 5\,k different pairs is generated, from which we compute the validation EER. This is used to select the best checkpoint during a training run as well as to tune hyperparameters. For evaluation we use the cleaned \textit{original} (vox1-o, 40 speakers, \textapprox37\,k trials), \textit{extended} (vox1-e, 1251 speakers, \textapprox580\,k trials) and \textit{hard} (vox1-h, 1190 speakers, \textapprox550\,k trials, equal nationality and sex) test sets from voxceleb 1\cite{Chung18b}. There is no speaker overlap between the voxceleb1 test sets and the voxceleb2 dev set. The experiments with the wav2vec2 network use the pre-trained weights\footnote{See \url{https://huggingface.co/facebook/wav2vec2-base}} on Librispeech \cite{panayotov2015librispeech} released on HuggingFace \cite{wolf-etal-2020-transformers} by Fairseq \cite{ott2019fairseq}.

\subsection{Computational budget and fair comparison}\label{budget}

We compare the performances of models under similar computational budgets. Each network is trained with a batch size of 3.2M audio samples \cite{NEURIPS2020_92d1e1eb} by randomly sampling 3 seconds from 66 different audio files. No data augmentation techniques are used. We train for 100k iterations, approximately 6 epochs, with Adam \cite{adamkingma14} and a OneCycle learning rate schedule \cite{smith2019super}. The maximum learning rate (LR) of the cycle is found by 1) performing an LR range test \cite{smith2017cyclical} with 5k iterations  and 2) tuning on a 7-sized grid centered around the LR with the steepest slope and bounded by the LRs where the loss respectively started and stopped decreasing. Models are evaluated with a cosine score between speaker embeddings lacking any further post-processing, except for the wav2vec2-bce variant which computes scores directly.

\subsection{Baseline systems}

We train two popular baselines models for speaker recognition, x-vector \cite{snyder2018x,speechbrain} and ECAPA-TDNN \cite{desplanques20_interspeech,speechbrain}, and compare them to the three wav2vec2 adaptions w2v2-ce, w2v2-aam and w2v2-bce. All five models have the computation budget described in subsection \ref{budget}. The X-vector and ECAPA-TDNN networks use 40-dimensional filterbanks as input. The w2v2-ce and w2v2-aam variants use mean\&std pooling which was chosen as it is also used in the x-vector network architecture. The w2v2-aam variant and ECAPA-TDNN use the AAM softmax loss with a scale of $30$ and a margin of $0.2$ in this and further experiments. The feature encoder part of the wav2vec2 architecture is frozen for the whole training procedure which corresponds to \cite{NEURIPS2020_92d1e1eb}.

\subsection{Variation in pooling}

Next we explore the different poolings methods proposed in Section \ref{sec:meth_pool}. This was carried out for the single-utterance classification variants: w2v2-ce and w2v2-aam. We use the same training settings as in the baseline comparison and only vary the pooling method. Note therefore that the learning rate was tuned to the ``mean\&std'' setup.

\begin{table}[t]
\centering
\caption{\fontsize{9}{11}\selectfont EER performance of standard filterbank-based speaker recognition networks as well as the fine-tuned wav2vec2 variations.  We ran $N=4$ runs to compute the standard deviations. Bold indicates best performance.}
\label{tab:eer_table}
\resizebox{\linewidth}{!}{%
\begin{tabular}{llllllllll}
\hline
\multicolumn{1}{r}{}        &  & \multicolumn{8}{c}{EER performance (mean, std in \%)}                                                 \\ \cline{3-10} 
\multicolumn{1}{c}{NETWORK} &  & \multicolumn{2}{l}{vox1-o} &      & \multicolumn{2}{l}{vox1-e} &      & \multicolumn{2}{l}{vox1-h} \\ \cline{1-1} \cline{3-4} \cline{6-7} \cline{9-10} 
x-vector \cite{snyder2018x}  &  & 5.22   & 0.12    &  &  5.60  & 0.05    &  & 8.75   & 0.05    \\
ECAPA-TDNN \cite{desplanques20_interspeech}    &  & \textbf{1.61}   & 0.03    &  & \textbf{1.69}    & 0.03    &  & \textbf{3.10}   & 0.05    \\ \hline
w2v2-ce   &  & 2.25 & 0.20 &  & 2.58 & 0.10 &  & 4.91 & 0.13 \\
w2v2-aam  &  & 1.91 & 0.12 &  & 2.22 & 0.04 &  & 4.33 & 0.08 \\
w2v2-bce  &  & 7.28    & 0.22    &  & 7.19    & 0.22    &  & 11.34    & 0.83    \\ \hline
\end{tabular}%
}
\end{table}

\subsection{Ablation study}

We perform several ablations on the best-performing wav2vec2 variant for speaker recognition. These ablations are trained with 100k iterations except in the experiments for the batch size. The first set of ablations study the effect of not freezing the feature extractor in the fine-tuning procedure as well as randomly initialising the whole network and thus not using any pre-trained weights. The second set of ablations explore the relative importance of the regularisation techniques in the network architecture. We first disable only LayerDrop, then sequentially disable dropout and the masking of certain frames as well. The third set simply studies the effect of increasing (with 50k iterations) or decreasing (with 200k iterations) the chosen batch size by a factor of 2. The last ablations involve the learning rate schedule. We first test two constant learning rates: $3\times \e-6$ is the LR where the loss started increasing in the LR range test and \e-5\ is the LR with the steepest decrease. The second schedule exponentially decays the LR from \e-5\ to $3\times \e-6$. The final schedule is the tri-stage learning rate schedule similar to \cite{NEURIPS2020_92d1e1eb} which includes a warm-up phase linearly increasing the LR from \e-7\ to \e-5\ in 10k iterations, a constant phase of 40k iterations, and an exponentially decreasing stage from \e-5\ to \e-7\ for the remaining iterations. 

\section{Results}
\label{sec:res}

\subsection{Baseline comparison}

The LR range test and grid-tuning approach found the following learning rates: \e-4\ for x-vector, \e-3\ for ECAPA-TDNN, $9\times\e-5$ for w2v2-ce, $5\times\e-5$ for w2v2-aam, and $3\times\e-5$ for w2v2-bce. Table \ref{tab:eer_table} shows four runs with these learning rates. We see that ECAPA-TDNN performs best on all test sets. The w2v2-aam network is the best performing wav2vec2 variant. Both w2v2-ce and w2v2-aam manage to improve on the x-vector architecture. Modeling speaker recognition as utterance-pair classification (w2v2-bce) performed worst.

\subsection{Pooling methods}

\begin{table}[t]
\fontsize{9}{11}\selectfont
\centering
\caption{\fontsize{9}{11}\selectfont The performance of different pooling strategies for the single utterance classification architectures on the extended voxceleb1 test set. Each method was trained with $N=3$ random seeds. The evaluation of random pooling was repeated four times for each run. 
}
\label{tab:pool}
\begin{tabular}{lllllll}
\hline
\multicolumn{1}{r}{}        &  & \multicolumn{5}{c}{EER on vox1-e (mean, std in \%)}                                   \\ \cline{3-7} 
\multicolumn{1}{c}{pooling} &  & \multicolumn{2}{c}{w2v2-ce} & \multicolumn{1}{c}{}     & \multicolumn{2}{c}{w2v2-aam} \\ \cline{1-1} \cline{3-4} \cline{6-7} 
max          &  & 4.79          & 0.55  &  & 2.27 & 0.04  \\
quantile     &  & 2.75          & 0.17  &  & 2.21 & 0.03  \\
mean         &  & 2.69          & 0.06  &  & 2.11 & 0.05  \\
mean\&std     &  & 2.60          & 0.08  &  & 2.18 & 0.03  \\
first        &  & 2.61          & 0.10  &  & 2.15 & 0.05  \\
first\&cls    &  & \textbf{2.52} & 0.11  &  & \textbf{2.06} & 0.03  \\
middle       &  & 2.55          & 0.07  &  & 2.18 & 0.03  \\
last         &  & 2.58          & 0.07  &  & 2.18 & 0.03  \\
random ($N=1$) &  & 2.56          & 0.002 &  & 2.40 & 0.002   \\
random ($N=3$) &  & 2.70          & 0.13  &  & 2.37 & 0.03   \\ \hline
\end{tabular}%
\end{table}

Table \ref{tab:pool} compares the 9 different pooling strategies with the w2v2-ce and w2v2-aam networks. We observe that first\&cls pooling performs best for both networks. The difference between first\&cls pooling and the other pooling methods is more pronounced for w2v2-aam than for w2v2-ce.
The low inter-model variance of random pooling shows that each wav2vec2 embedding is a stand-alone speaker embedding. Not shown, using random pooling in the evaluation for networks which were trained with first\&cls pooling degrades the EER with $0.2\%$ points. Moreover, creating ensembles out of different embeddings in the output sequence did not improve performance. This suggests that the transformer layer processes each embedding in the sequence similarly. 

\subsection{Ablation study}
Table~\ref{tab:ablation} shows the results of the ablation study.
We see that freezing the feature extractor leads to worse performance but is more stable across runs with different seeds. We also note a large degradation in performance when initializing with random weights instead of the pre-trained weights. The regularisation settings for fine-tuning on speech recognition are also beneficial for fine-tuning on speaker recognition. Increasing the batch size beyond 3.2M audio samples does not increase performance, although it does decrease the variance slightly. Using a learning rate schedule with a warm-up phase, such as the tri-stage or OneCycle schedule, is critical for stable and good performance. 

\begin{table}[t]
\fontsize{9}{11}\selectfont
\centering
\caption{\fontsize{9}{11}\selectfont EER performance under varying ablated configurations for the w2v2-aam network variant. Each configuration was run with $N=3$ random seeds. One run with exponential decay diverged and we therefore show $N=2$ results for that row.}
\label{tab:ablation}
\resizebox{\linewidth}{!}{%
\begin{tabular}{llccc}
\hline
\multicolumn{1}{c}{}         &  & \multicolumn{3}{c}{EER on vox-e}                                                   \\ \cline{3-5} 
\multicolumn{1}{c}{ablation} &  & \multicolumn{1}{l}{mean(\%)} & \multicolumn{1}{l}{} & \multicolumn{1}{l}{std (\%)} \\ \cline{1-1} \cline{3-3} \cline{5-5} 
default w2v2-aam first\&cls setup            &  & 2.06 & \multicolumn{1}{l}{} & 0.03 \\ \hline
unfrozen feature extractor   &  & 1.88 &                      & 0.08 \\
\& no pre-trained weights    &  & 5.08 &                      & 0.08 \\ \hline
no layerdrop ($p=0$)         &  & 2.45 &                      & 0.05 \\
\& no dropout ($p=0$)        &  & 2.39 &                      & 0.04 \\
\&\& no time masking ($p=0$) &  & 2.39 &                      & 0.09 \\ \hline
batch size 32                &  & 2.12 &                      & 0.10 \\
batch size 128               &  & 2.05 &                      & 0.01 \\ \hline
constant (1e-5)              &  & 50.0 &                      & 0.00 \\
constant (3e-6)              &  & 3.71 &                      & 0.08 \\
exponential decay ($N=2$)      &  & 2.42 &                      & 0.07 \\
tri-stage                    &  & 1.97 &                      & 0.04 \\ \hline
\end{tabular}%
}
\end{table}

\section{Conclusion and future work}
\label{sec:concl}

We have shown that the wav2vec2 framework can be successfully adapted to the speaker recognition task and that the pre-trained weights used for fine-tuning on speech recognition are also useful for fine-tuning on speaker recognition. Good results with first\&cls pooling indicate that including a class token in the pre-training procedure is promising future work. Modeling speaker recognition as a paired-utterance classification problem did not perform well. Future work in that direction might involve limiting the attention mechanism to the opposite utterance to simplify the learning task.



\ninept

\bibliographystyle{IEEEbib}
\bibliography{strings,refs}

\end{document}